\documentclass[fleqn,10pt]{wlscirep}

\newcommand{\Tr}{\operatorname{tr}}
\def\one{\leavevmode\hbox{\small1\normalsize\kern-.33em1}}
\newcommand{\ketbra}[2]{\left|#1\middle\rangle\middle\langle#2\right|}
\newcommand{\ZZ}{\mathbb{Z}}

\title{EPR Steering inequalities with Communication Assistance}

\author[1]{S\'andor Nagy}
\author[2]{Tam\'as V\'ertesi}
\affil[1]{Department of Theoretical Physics, University of Debrecen, H-4010 Debrecen, P.O. Box 5, Hungary}
\affil[2]{Institute for Nuclear Research, Hungarian Academy of Sciences, H-4001 Debrecen, P.O. Box 51, Hungary}

\begin{abstract}
In this paper, we investigate the communication cost of reproducing Einstein-Podolsky-Rosen (EPR) steering correlations arising from bipartite quantum systems. We characterize the set of bipartite quantum states which admits a local hidden state model augmented with $c$ bits of classical communication from an untrusted party (Alice) to a trusted party (Bob). In case of one bit of information ($c=1$), we show that this set has a nontrivial intersection with the sets admitting a local hidden state and a local hidden variables model for projective measurements. On the other hand, we find that an infinite amount of classical communication is required from an untrusted Alice to a trusted Bob to simulate the EPR steering correlations produced by a two-qubit maximally entangled state. It is conjectured that a state-of-the-art quantum experiment would be able to falsify two bits of communication this way.
\end{abstract}

\begin{document}

\flushbottom
\maketitle

\thispagestyle{empty}

\section*{Introduction}

Quantum entanglement is a remarkable phenomenon that has no counterpart in classical physics~\cite{entanglement1,entanglement2}.
Beyond its fundamental importance, it is a crucial resource in quantum information
and quantum computing~\cite{nielsenchuang}. Entanglement gives rise to the phenomenon of Bell nonlocality \cite{Bell,NL-review}, which lies at the heart of device-independent quantum information processing \cite{DIscarani}. Such device-independent protocols are greatly immune against errors which are due to deviations of the ideal description of the setup from the actual physical implementation.

There is an intermediate form of non-separability between entanglement and nonlocality linked to the phenomenon of
Einstein-Podolsky-Rosen (EPR) steering \cite{sch35}, which was put on a firm basis recently by Wiseman, Doherty and Jones \cite{steering1,steering2} by introducing an information task for arbitrary quantum systems. Since then, both the detection~\cite{detEPR1,detEPR2,detEPR3,detEPR4} and quantification~\cite{resource11,resource12,resource2,resource3,paulmigueldani,pusey,marco} of EPR steering have been thoroughly investigated with interesting applications in quantum information~\cite{appEPR1,appEPR2,multist1} and recent experimental tests~\cite{expEPR1,expEPR2,expEPR3,expEPR4,steeringNP}. More recent experiments have addressed multipartite quantum steering~\cite{multist2} and one-way steering~\cite{expEPR5,expEPR6}.

\begin{figure}[htb!]
\centering
\includegraphics[width=0.4\columnwidth,trim={0cm 0cm 11cm 0cm},clip]{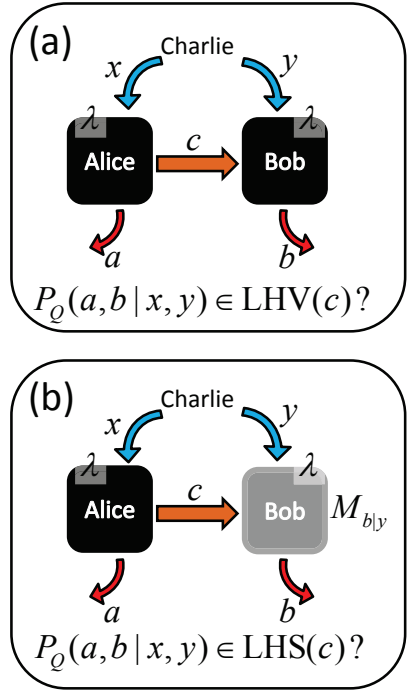}
\caption{The setup for simulating (a) Bell nonlocal and (b) EPR steering correlations with local models using auxiliary communication. In (a) the simulation protocol is as follows. The two parties distribute shared randomness $\lambda$.
Charlie sends settings $(x,y)$ to the two parties. After obtaining the settings, Alice is allowed to communicate to Bob a classical message consisting of $c$ bits. Finally, Alice and Bob give outputs $a$ and $b$ as a function of available information for each party. The (b) protocol is similar to (a) with the difference that Charlie fully trusts Bob, hence, we can assume that Bob performs a given set of quantum measurements $\{M_{b|y}\}_{b,y}$ on $\sigma_{\lambda}$.}
\label{fig_setup}
\end{figure}

Quantum correlations can be phrased in terms of an information task wherein a referee, say Charlie, wants to verify that two parties, called Alice and Bob, share an entangled state (see Fig.~\ref{fig_setup} displaying the setup). In the preparation stage of the protocol, Alice and Bob share a number of copies of a bipartite state $\rho$, and for each of those states Charlie asks them to perform one of a number of measurements chosen by Charlie at random. Alice's and Bob's measurements are denoted by $M_{a|x}$ and $M_{b|y}$, respectively, where $x$ and $y$ denote the choice of measurements and $a$ and $b$ their corresponding outputs. By repeating the procedure many times, they form the joint probability distribution $P_Q(ab|xy)$, which is given by
\begin{equation}
\label{p_abxy}
P_Q(ab|xy)=\Tr(\rho M_{a|x}\otimes M_{b|y}).
\end{equation}
That is, the object of our study is the probability distribution of the outputs of the two parties dependent on each party's input (i.e. choice of measurement settings). Throughout we will assume that measurements are projective ones, that is, $M_{a|x}^2=M_{a|x}$ and $M_{b|y}^2=M_{b|y}$. Note that for two-outcome settings (which is our main concern) this is not a limitation \cite{cleveetal}.

Basically, there are three options to certify entanglement depending on the number of trusted parties participating in the protocol. Charlie trusts both Alice and Bob (and their apparatuses). Charlie trusts (say) Bob, but not Alice. Finally, Charlie trusts neither Alice nor Bob.

In the latter case of no trust at all (i.e. the Bell nonlocality scenario), we say that a quantum state $\rho$ is Bell local or equivalently admits a local hidden variables (LHV) model (for projective measurements), when the statistics $P_Q(ab|xy)$ originating from arbitrary local (projective) measurements $M_{a|x}$ and $M_{b|y}$ in (\ref{p_abxy}) can be reproduced by a distribution of the form
\begin{equation}
\label{model}
P(ab|xy)=\sum_{\lambda}P(\lambda)P_{\lambda}(a|x)P_{\lambda}(b|y),
\end{equation}

\noindent where $\lambda$ is some shared classical random variable distributed according to the density $P(\lambda)$, and $P_{\lambda}(a|x)$ and $P_{\lambda}(b|y)$ are arbitrary local response functions of Alice and Bob, respectively. In that case, the distribution $P_Q(ab|xy)$ cannot violate any Bell inequality. Conversely, if the distribution $P_Q(ab|xy)$ cannot be written in the form~(\ref{model}), it violates a Bell inequality. This implies that some form of extra communication is required between Alice and Bob in order to reproduce the statistics $P_Q(ab|xy)$.

On the other hand, in case of partial trust (i.e. an EPR steering scenario), we obtain the data $P_Q(ab|xy)$ with an additional knowledge that Bob's system is well-characterized. That is, Charlie trusts Bob's measurements $\{M_{b|y}\}_{b,y}$. In that case, the state $\rho$ shared by Alice and Bob is said to be unsteerable or equivalently to admit a local hidden state (LHS) model, when the statistics $P_Q(ab|xy)$ can be reproduced by a distribution of the form (\ref{model}), where now
\begin{equation}
\label{Pby}
P_{\lambda}(b|y)=\Tr(\sigma_{\lambda}M_{b|y}).
\end{equation}
In that case, the distribution $P_Q(ab|xy)$ cannot violate the so-called steering inequalities. Conversely, if the distribution $P_Q(ab|xy)$ cannot be written in the form~(\ref{model}) with the above restriction~(\ref{Pby}) on $P_{\lambda}(b|y)$, it violates a steering inequality. Again, this means that some communication has to be taken place between Alice and Bob to reproduce the obtained statistics $P_Q(ab|xy)$.

Finally, in the case that Charlie trusts both Alice's and Bob's measurement devices, the LHS model above becomes a quantum separable (QS) model, where there exist  local density operators $\sigma_{\lambda}^A$ and $\sigma_{\lambda}^B$ such that the response functions of Alice and Bob in the formula~(\ref{model}) are given respectively by $P_{\lambda}(a|x)=\Tr(\sigma_{\lambda}^A M_{a|x})$ and $P_{\lambda}(b|y)=\Tr(\sigma_{\lambda}^B M_{b|y})$. Failure of satisfying this model implies entanglement. This again can be detected through the violation of certain inequalities which are conventionally called entanglement witnesses.

However, observing either a violation of a LHV model in the Bell nonlocality scenario, or violation of a LHS model in the EPR steering scenario, or violation of a QS model in the entanglement scenario will not quantify the amount of communication beyond the fact that some communication was indeed required. In case of Bell nonlocality, where no trust is assumed in either devices, Bacon and Toner provided a general framework for a measure of nonlocality by allowing the parties to communicate some bits of information after selecting the measurement settings \cite{bacontoner}. They in particular proved that correlations produced by projective measurements on the two-qubit singlet state can be simulated with a LHV model augmented by a single bit of communication \cite{tonerbacon}.

In this paper, we pose an analogous question in the EPR steering scenario. Our aim is to quantify the correlations arising from quantum (projective) measurements conducted by Alice on her share of an entangled particle. To do so, we allow some amount of classical communication from the untrusted party Alice to a trusted party Bob.

The structure of the paper is as follows. We first present the Bell nonlocality setup by Bacon and Toner~\cite{bacontoner}. Then we translate this setup to the EPR-steering scenario. In particular, we develop a computational framework to decide if an EPR correlation produced by quantum theory can be simulated by a LHS model plus exchanging a number of bits of communication. We provide an efficient code based on semidefinite programming (SDP)~\cite{sdp} to solve such a membership problem. This allows us to explore the shape of the set of two-qubit states admitting a LHS model augmented with one bit of communication, a set we denote by $LHS(1)$. Specifically, we prove that the set $LHS(1)$ is strictly larger than the set of states admitting a LHS model (for projective measurements). On the other hand, we conduct an extensive numerical search which indicates that there exist two-qubit quantum states admitting a LHV model (assuming projective measurements), which nevertheless cannot be described by a LHS model assisted with 1 bit of classical communication. Finally, we show that an infinite amount of classical communication is required from an untrusted Alice to a trusted Bob to simulate the statistics of any bipartite pure entangled state in this scenario.

\section*{Results}

\subsection*{Bell nonlocality with communication}

Bacon and Toner use the following classical protocol to simulate a Bell scenario~\cite{bacontoner} (see also Ref.~\cite{maxwell} for more recent results). Let us consider $c=\log_2 d$ bits of classical communication sent from Alice's device to Bob's device. Upon receiving input $x$, Alice's device sends communication $c$ and outputs $a$ with probability $P(a,c|x)$. Upon receiving input $y$ and communication $c$, Bob's device outputs $b$ with probability $P(b|y)$. We thus have that
\begin{equation}
\label{lhvc}
P(ab|xy)=\sum_{c=1}^d P(a,c|x)P(b|y,c).
\end{equation}

\noindent In the simulation protocol, we also would like to take into account the possibility that Alice and Bob's devices are correlated due to a common random variable $\lambda$, which was prepared and distributed between the parties before receiving inputs $x$ and $y$ from Charlie. In that case, the set of admissible distributions is formed by all convex combination of strategies labeled by $\lambda$:

\begin{equation}
\label{lhvclambda}
P(ab|xy)=\sum_{\lambda}P(\lambda)\sum_{c=1}^d P_{\lambda}(a,c|x)P_{\lambda}(b|y,c),
\end{equation}
where each $P(\lambda)\ge0$ and $\sum_{\lambda}P(\lambda)=1$. Bacon and Toner~\cite{bacontoner} quantify the amount of resource by the number of bits of communication $c$ to match $P_Q(ab|xy)$ with the distribution $P(ab|xy)$ in Eq.~(\ref{lhvclambda}). They prove that $c=1$ bit of communication assistance (i.e. sending a message with $d=2$ levels) is enough for Alice and Bob to reproduce any correlations $P_Q(ab|xy)$ by measuring arbitrary projective measurements on a maximally entangled two-qubit state~\cite{tonerbacon}. Fig~\ref{fig_setup}.(a) displays this setup.

\subsection*{EPR steering with communication}

We ask the analogous question what happens if (unlike in case of the Bell nonlocality scenario) Charlie completely trusts Bob's measurement device. This is the framework of EPR steering. In this case, we obtain the data $P_Q(ab|xy)$ with an additional knowledge that Bob's equipment is well-characterized. Hence our model is the same as (\ref{lhvclambda}), but with the constraint that
\begin{equation}
P_{\lambda}(b|y,c)=\Tr(\sigma_{\lambda,c}M_{b|y}),
\end{equation}
where Bob's measurements $\{M_{b|y}\}_{b,y}$ are trusted by Charlie and the states $\sigma_{\lambda,c}$ are of unit trace and positive. Note, however, that in this case Bob's measurements will not depend on the communication $c$, since they can be considered as supplied by Charlie. Please see Fig.~\ref{fig_setup}.(b), which shows this setup. Hence we get
\begin{equation}
\label{lhsclambda}
P(ab|xy)=\sum_{\lambda}P(\lambda)\sum_{c=1}^d D_{\lambda}(a,c|x)\Tr(\sigma_{\lambda,c}M_{b|y}),
\end{equation}
where $D_{\lambda}(a,c|x)$ are some deterministic strategies labeled by $\lambda$ taking values 0 or 1. Note that we used the fact that unshared randomness of $P_{\lambda}(a,c|x)$ can always be considered as part of shared randomness (represented by $\lambda$) and hence can be absorbed into it. For instance, if Alice performs $m$ measurements ($x=1,\ldots,m$) with two outcomes each ($a=1,2$), and the communicated message is one bit ($c=1,2$), each deterministic strategy $D_{\lambda}$ can be considered as an $m$-component vector $\vec v = (v_1,\ldots,v_m)$  with four possible entries $(1,1)$, $(1,2)$, $(2,1)$, and $(2,2)$ standing for the values $(a,c)$. Hence, in this case there are $4^m$ different deterministic strategies for Alice.

In fact, given the statistics $P_Q(ab|xy)$ and a set of measurement operators $\{M_{b|y}\}_{b,y}$ in Eq.~(\ref{lhsclambda}), this is a feasibility problem to check if a LHS model augmented with $c$ bits of communication can reproduce the distribution $P_Q(ab|xy)$. If so, the underlying state of the distribution $P_Q(ab|xy)$ in (\ref{p_abxy}) can be reproduced with a LHS model with some extra communication $c$. Then, by definition, $\rho$ is within the set $LHS(c)$.

We can write this feasibility problem as an SDP code. To this end, we define the sub-normalized states $\tilde\sigma_{\lambda,c}=P(\lambda)\sigma_{\lambda,c}$ for all $(\lambda,c)$, which satisfy $\Tr\tilde\sigma_{\lambda,c}=\Tr\tilde\sigma_{\lambda,c'}$ for all different pairs $(c,c')$ and $\sum_{\lambda}\Tr\tilde\sigma_{\lambda,c}=1$ for all $c=1,\ldots,d$.
Then, we have to solve the following feasibility SDP program:
\begin{equation} \label{feasibsdp}
\begin{aligned}
\text{find}\quad &\left\{ \sigma_{\lambda,c}\right\} \\
\text{subject to}\quad &P_Q(ab|xy)=\sum_{\lambda}\sum_{c=1}^d D_{\lambda}(a,c|x)\Tr(\tilde\sigma_{\lambda,c}M_{b|y})  &\forall a,b,x,y \\
\quad &\Tr \tilde\sigma_{\lambda,c} = \Tr \tilde\sigma_{\lambda,c'} &\forall \lambda,c\neq c'\\
\quad &\sum_{\lambda}\Tr\tilde\sigma_{\lambda,c}=1 &\forall c\\
\quad &\tilde\sigma_{\lambda,c}\ge 0 &\forall \lambda,c
\end{aligned}
\end{equation}
where the data $P_Q(ab|xy)$ is arising from formula~(\ref{p_abxy}) and $M_{b|y}$ are fixed measurements of Bob.
We can in fact simplify a bit the above code by eliminating Bob's measurements $M_{b|y}$.
To this end, let us define the conditional (unnormalized) states prepared by Alice on Bob's subsystem by
\begin{equation}
\label{assem}
\sigma_{a|x} = \Tr_A(\rho M_{a|x}\otimes\one).
\end{equation}
This set of states is called assemblage~\cite{pusey,paulmigueldani} and captures the whole physics of an EPR steering scenario. With this assemblage, we have $P_Q(ab|xy)=\Tr(\sigma_{a|x}M_{b|y})$. Let us assume that Bob's measurements $M_{b|y}$ form a complete basis of Bob's Hilbert space. In case of a two-dimensional Hilbert space, this can be the three Pauli matrices,
$M_{b|y}=(\one+(-1)^b \hat\sigma_y)/2$, where $\hat\sigma_y$, $y=1,2,3$ denote the Pauli operators and the two possible outcomes are denoted by $b=0,1$.
Then, the SDP code simplifies to
\begin{equation} \label{feasibsdp2}
\begin{aligned}
\text{find}\quad &\left\{ \sigma_{\lambda,c}\right\} \\
\text{subject to}\quad &\sigma_{a|x}=\sum_{\lambda}\sum_{c=1}^d D_{\lambda}(a,c|x)\tilde\sigma_{\lambda,c} &\forall a,x \\
\quad &\Tr \tilde\sigma_{\lambda,c} = \Tr \tilde\sigma_{\lambda,c'} &\forall \lambda,c\neq c'\\
\quad &\sum_{\lambda}\Tr\tilde\sigma_{\lambda,c}=1 &\forall c\\
\quad &\tilde\sigma_{\lambda,c}\ge 0 &\forall \lambda,c
\end{aligned}
\end{equation}
It is worth noting that the above code simplifies to the one derived in Ref.~\cite{paulmigueldani} in case of $c=0$ bit of communication (that is, $d=1$).

\subsection*{Exploring the shape of the $LHS(1)$ set of states}

\begin{figure}[htb!]
\centering
\includegraphics[width=0.5\columnwidth,trim={2cm 5cm 2cm 3cm},clip]{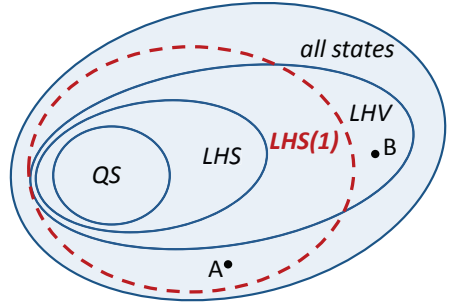}
\caption{Schematic view of the different set of states. All depicted sets are convex. The smallest set corresponds to quantum separable (QS) states, the largest set contains all states. States which have a LHV model (i.e. Bell local) are in between these sets. States which have a LHS model (i.e. unsteerable) are sandwiched between the $LHV$ and $QS$ sets. The new set (whose boundary is drawn by a dashed line) is termed as $LHS(1)$ and it has a nontrivial intersection with the $LHS$ and $LHV$ sets. In this paper, we prove the existence of point~A and conjecture supported by extensive numerical calculations the existence of point~B.}
\label{fig_sets}
\end{figure}

In this subsection, we explore the shape of the $LHS(c)$ set, where we set $c=1$ bit. Specifically, we ask how it fits into the set of bipartite states which admit a LHS or LHV model. Let us note that the new set $LHS(1)$ is also convex by construction. We find a nontrivial structure of this new set. To this end, we investigate special one-parameter slices of the full two-qubit state space. In particular, we choose two special one-parameter families, the Werner states of two-qubits and another family, which coincides with the two-qubit reduced state of the $n$-qubit $W_n$ state \cite{dur} for parameters $p=2/n$. The obtained results suggest that the $LHS(1)$ set of states has a nontrivial shape as depicted in the schematic picture of Fig.~\ref{fig_sets}.

We use the SDP code~(\ref{feasibsdp2}) to test one-parameter families of two-qubit quantum states. Namely, let us write the state as a mixture of a pure entangled state and a noisy part parameterized by the weight $v$:
\begin{equation}
\label{onepara}
\rho(v) = v|\psi\rangle\langle\psi|+(1-v)\rho_{noise},
\end{equation}
where $\rho_{noise}$ is some fixed separable state and $|\psi\rangle$ is any two-qubit entangled pure state. A small variation of the semi-definite program~(\ref{feasibsdp2}) (please see Methods section for the actual code) gives us an efficient method to place an upper bound on $v_{crit}$, where $v_{crit}$ denotes the boundary of states admitting a LHS(1) model. Such numerical computations, as well as all subsequent ones presented in this paper, were carried out using the Matlab packages YALMIP \cite{yalmip} and the SDP solver SeDuMi \cite{sedumi}.

Then, using a heuristic search (e.g., we used an Amoeba routine~\cite{NM}), we lower the value of this upper bound on $v_{crit}$ by varying the set of measurements $\{M_{a|x}\}_{a,x}$. This way, we get better and better upper bounds to the true value of $v_{crit}$ by minimizing the parameters entering Alice's set of measurements $\{M_{a|x}\}_{a,x}$. We remark that due to the heuristic nature of the search, the program may not provide us a global minimum for $v_{crit}$, however, for reasonable number of settings (say, $m_A\le 6$) and a fair number of independent iterations, the obtained bound to our experience is quite reliable.

\begin{figure}[htb!]
\centering
\includegraphics[width=0.6\columnwidth,trim={1cm 10cm 3cm 1cm},clip]{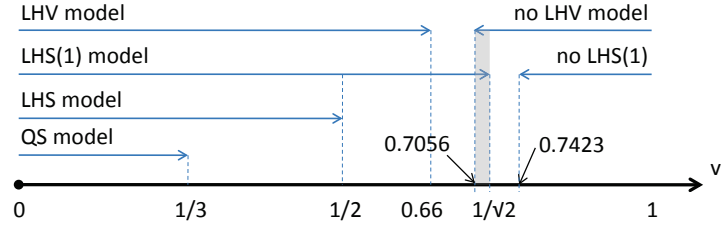}
\caption{Regions of the parameter $v$ in which the two-qubit Werner state is quantum separable, admits an LHS, LHS(1), and LHV models. It shows the shaded interval $[0.7056,1/\sqrt 2\simeq0.7071]$, where the state has a LHS(1) model, nevertheless it is nonlocal. We note that the values $1/3$ and $1/2$ corresponding to the respective QS and LHS models are tight. That is, any $v$ larger than these values results in failure of these models. However, according to the figure, this is not the case for the LHS(1) and LHV models and there arises a gap between the best upper and lower bounds on the critical value of $v$.}
\label{fig_werner}
\end{figure}

We first consider the Werner state of two qubits~\cite{Werner}, which is given by
\begin{equation}
\label{werner}
\rho_{W}(v) = v|\psi_-\rangle\langle\psi_-| + (1-v)\frac{\one\otimes\one}{4},
\end{equation}
where $|\psi_-\rangle=(|01\rangle-|10\rangle)/\sqrt 2$ is the singlet state and $v$ is the visibility. The Werner state is separable up to $v=1/3$ and exhibits a LHS model up to $v=1/2$~\cite{steering1,steering2,steeringNP}. These models are tight, hence $v>1/3$ implies entanglement, whereas $v>1/2$ implies violation of EPR steering inequalities.

Concerning the LHS(1) model, we find the following results. First, in the Methods section a LHS(1) model is provided up to visibility $v=1/\sqrt 2\simeq0.7071$ of the Werner states. On the other hand, Amoeba optimization provides us with a steering inequality for $m=4$ settings which is violated above the parameter $v=0.7842$. Also, by setting Alice's 12 measurements to point toward the vertices of an icosahedron on the Bloch sphere, we get a more powerful steering inequality which is violated above $v=0.7423$ (note that due to reflection symmetry of the icosahedron, it is enough to consider $m=6$ vertices in the actual code). Therefore, there is no LHS(1) model below $v=0.7423$, as depicted in Fig.~\ref{fig_werner}. We state it as an open problem what the exact value of the critical $v$ above which no LHS(1) model exists if $m$ goes to infinity. On the other hand, there is a 465 setting Bell inequality \cite{vertesiKG3,Hua}, which is violated by a Werner state above $v=0.7056$, which implies that there exists no LHV model for the Werner state for $v>0.7056$. Again, this bound is shown in Fig.~\ref{fig_werner}. The above bounds entail that the $LHS(1)$ set has portions outside the $LHV$ (i.e. Bell local) set of states. This is the shaded region depicted in Fig.~\ref{fig_werner} and proves in turn the existence of point~A in the schematic Fig.~\ref{fig_sets} (i.e., a state which is nonlocal and admits a LHS(1) model). Since the set of states admitting a LHV model is a strict superset of the states admitting a LHS model~\cite{toner1,toner2}, it follows that the $LHS(1)$ set is strictly different from the $LHS$ set. We provide an alternative proof of this fact in the Methods section.

Note that the hierarchy $QS\subseteq LHS\subseteq LHV$ of the sets is implied by the definitions. Moreover, it is known due to the works of Refs.~\cite{Werner,toner1} that the above relations are strict, that is, we have $QS\subsetneq LHS\subsetneq LHV$. It is interesting to ask if the same hierarchy applies in the presence of a fixed amount of communication (say 1 bit). Indeed, implied by the definition of these sets, we have $QS(c)\subseteq LHS(c)\subseteq LHV(c)$ for any $c$ bits. We now show that the inclusion relations are strict, that is, $QS(c)\subsetneq LHS(c)\subsetneq LHV(c)$ for any finite number of $c$ bits. The first strict inclusion relation comes from the fact that there is no QS($\infty$) model (and consequently no QS($c$) model for any $c$ as well) for the Werner state for $v>1/2$. A sketch of this proof is deferred to the Methods section. Recalling that the Werner state admits a LHS(1) model (and consequently a LHS($c$) model for $c\ge 1$) up to $v=1/\sqrt 2$, it follows the strict relation $QS(c)\subsetneq LHS(c)$. The second strict relation $LHS(c)\subsetneq LHV(c)$ in case of $c=1$ comes from the fact that the Werner state for parameter $v=1$ admits a LHV(1) model due to the model of Toner and Bacon~\cite{tonerbacon} and on the other hand there is no LHS(1) model above $v=0.7423$ due to our result. Furthermore, in the following section we prove that no LHS($c$) model with finite $c$ exists for the two-qubit maximally entangled state (i.e., for the Werner state with v=1). Then we have $LHS(c)\subsetneq LHV(1)$ for any finite $c$ and the second strict relation $LHS(c)\subsetneq LHV(c)$ follows for any finite $c\ge 1$.

The other family of states to be investigated looks as follows:
\begin{equation}
\label{RW}
\rho_{R}(p) = p|\psi_+\rangle\langle\psi_+| + (1-p)|0\rangle\langle0|\otimes|0\rangle\langle0|,
\end{equation}
where $|\psi_+\rangle=(|01\rangle+|10\rangle)/\sqrt 2$. Notice that this state is the two-qubit reduced state of the $n$-qubit $W_n$ state \cite{dur} for $p=2/n$. We note that for this particular $p=2/n$, the state is $(n-1)$-symmetric extendable \cite{doherty}, hence there is a LHV model (and therefore also a LHS model) for $n-1$ settings (with arbitrary number of outcomes). The LHS bound seems to be tight, as we could recover the bound of $v=2/n$ up to numerical precision for $n\le6$ settings using the SDP method developed in Ref.~\cite{paulmigueldani} (please see second column of Table~\ref{w00}). This correspondence suggests that there is no LHS model for any finite $p>0$ if the number of settings is large enough. Using our numerical search described in the Methods section, we find the threshold values $p$ regarding the LHS(1) model in the third column of Table~\ref{w00}.

On the other hand, we conjecture that the local bound $p_{\textrm{LHV}}$ is $1/\sqrt 2\simeq0.7071$. Our conjecture is based on a linear programming approach combined with a heuristic search over the measurement angles \cite{LP} of Alice and Bob to get an upper bound on $p_{\textrm{LHV}}$ for a given number of measurement settings. For two settings per party ($m=2$), we have the (analytical) upper bound of $1/\sqrt 2$ on $p_{\textrm{LHV}}$. However, by moving up to $m=8$ settings per party, using numerical computations, this upper bound value did not become lower. Note that due to the heuristic nature of the search we cannot guarantee that $1/\sqrt 2$ is the optimal value. Though, at this level of complexity we are fairly confident about the validity of this threshold value. Moreover, we conjecture that this bound cannot be beaten beyond $m=8$ settings as well. Similar conclusion was drawn by Amirtham~\cite{diploma}. The above results (modulo our conjecture) indicate a point~B in Fig.~\ref{fig_sets}, displaying portions of the $LHV$ set lying outside the $LHS(1)$ set.

\begin{table}[t]
\centering
\vskip 0.2truecm
\begin{tabular}{c|c c}
\hline
$\text{\# settings}$&$p_{\textrm{LHS}}$&$p_{\textrm{LHS(1)}}$\\
\hline
2&$0.6667$&1\\
3&$0.5000$&0.8084\\
4&$0.4000$&0.7099\\
5&$0.3333$&0.6278\\
6&$0.2857$&0.5677\\
\hline
\end{tabular}
\caption{Table for certain critical parameters $p$ for the one-parameter family of two-qubit states given by formula~(\ref{RW}). The leftmost column stands for the number of settings, whereas the next two columns show (upper bounds to) the critical $p$ value with respect to number of settings for a LHS model and a LHS(1) model, respectively.} 
\label{w00}
\end{table}

\subsection*{Steering-like inequalities with any finite number of communication}

In this subsection, we go beyond the case of one bit of communication (i.e., $c=1$). To this end, we construct a steering inequality with $c=\log_2(d)$ number of bits of communication, which can be violated by a 2-qubit maximally entangled state for any finite $d$ if the number of settings $m$ for Alice is large enough. Violation implies that there is no LHS(c) model for any finite number of $c$ bits for a 2-qubit maximally entangled state. Combing this result with a recent work of Ref.~\cite{marco} entails that the same applies to any pure bipartite entangled state. More details about equivalence of states with respect to LHS(c) models are found in the Methods section.

Let us also remark that there is an interesting nested feature of the sets $LHS(c)$, namely they satisfy $LHS(c-1)\subseteq LHS(c)$ for any $c\ge 1$ implied by the definition (where we identified $LHS(0)\equiv LHS$). Moreover, in case of $c=1$, we have just shown that the inclusion relation is strict, that is $LHS\subsetneq LHS(1)$. It can be shown that in case of $c\rightarrow\infty$ all states are recovered, that is, the set $LHS(c\rightarrow\infty)$ approaches the set of all quantum states. We conjecture and state it as an open problem whether $LHS(c-1)\subsetneq LHS(c)$ holds true in case of any finite $c\ge 1$.

A steering inequality with communication assistance is a linear functional of the joint probabilities $P(ab|xy)$,
\begin{equation}
\label{sineq}
S\equiv \sum_{a,b,x,y}\alpha_{a,b,x,y}P(ab|xy)\le L_c,
\end{equation}
where the bound $L_c$ holds for any statistics $P(ab|xy)$ of the form~(\ref{lhvclambda}) arising from a LHS(c) model. Note that in the absence of communication ($c=0$), we return to the standard steering inequalities. Hence, if a $P_Q$ distribution of the form~(\ref{p_abxy}) violates bound $L_c$ in (\ref{sineq}), it implies that the underlying state of the probability distribution $P_Q$ lies outside the $LHS(c)$ set.

Let us consider a steering-like inequality augmented with $c$ bits of communication involving $m$ binary outcome settings both on Alice and Bob's side~\cite{steeringNP},
\begin{equation}
\label{naivesteering}
S_m = \frac{1}{m}\sum_{x,y=1}^m\delta_{x,y}E_{x,y}\le L_c
\end{equation}
\noindent where $\delta$ is the Kronecker delta function, $E_{x,y}=P(00|xy)+P(11|xy)-P(01|xy)-P(10|xy)$ is the expectation value involving Alice's $x$ and Bob's $y$ dichotomic measurements. Let Bob's observables $B_y=\vec u_y\cdot\vec\sigma$ point toward fixed dimensions $\vec u_y$. In particular, let us arrange Bob's observables to lie on the x-z plane of the Bloch sphere, $\vec u_y = (\cos(\pi y/m),0,\sin(\pi y/m))$ for settings $y=1,\ldots,m$.

According to model~(\ref{lhsclambda}), in case of communicating $c=\log_2 d$ bits (i.e., a $d$ level classical message is sent from Alice to Bob), the $L_c$ value corresponding to the LHS(c) limit is defined by maximizing the following expression
\begin{equation}
S_m = \frac{\sum_{x=1}^m E_{x,x}}{m} = \frac{\sum_{x=1}^m\langle A_x \vec u_x\cdot\vec\sigma\rangle_{\rho_{r(x)}}}{m}
\end{equation}
over all possible sign functions $A_x=\pm 1$, $x=1,\ldots,m$ and qubit states $\rho_{r(x)}$ with all possible $r(x)$ function $\ZZ_m \rightarrow \ZZ_d$, where $\vec u_x = (\cos(\pi x/m),0,\sin(\pi x/m))$.

Let us now choose the particular case of sending $c=\log_2(m-1)$ bits (i.e. Alice communicates a $d=m-1$ level message to Bob).
Then, $L_c$ is given by maximizing
\begin{equation}
S_m = \frac{m-2 + \langle A_{x}\vec u_{x}\cdot\vec\sigma + A_{x'}\vec u_{x'}\cdot\vec\sigma\rangle_{\rho}}{m},
\end{equation}
where $\rho$ is any single qubit state and $x\neq x'\in\{1,\ldots,m\}$. Let us choose the state $\rho$ optimally as the eigenstate of
\begin{equation}
\vec u_{x}\cdot\vec\sigma + \vec u_{x'}\cdot\vec\sigma,
\end{equation}
which reduces to maximizing
\begin{equation}
S_m = \frac{m-2 + |A_{x}\vec u_{x}+A_{x'}\vec u_{x'}|}{m}.
\end{equation}
This expression is maximized by e.g. choosing $x=1$, $x'=2$ and $A_{1}=A_{2}=+1$, resulting in the LHS(c) maximum:
\begin{equation}
\label{Lc}
L_c = 1 - 2\frac{1-\cos(\pi/2m)}{m}
\end{equation}
corresponding to $c=\log_2(m-1)$ bits of communication.

As we can see, the value $L_c$ is strictly smaller than 1 for any $m$. On the other hand, quantum mechanics allows us to obtain the algebraic bound of 1 in the left-hand side of the inequality~(\ref{naivesteering}). The quantum strategy comprises a maximally entangled state $|\psi_-\rangle$ and Alice's measurements $\vec v_x = -\vec u_x$, for $x=1,\ldots,m$ whereby we get the perfect correlation $E_{xx}=1$ for all $x$.

Therefore, we have an example, where we are unable to simulate quantum strategies by augmenting the LHS model with any finite number of bits of communication $c=\log_2(m-1)$ from Alice to Bob. Note, however, that as $m$ goes to infinity the $L_c$ value becomes close to 1, resulting in a very poor noise resistance. We pose it as an intriguing problem to construct more powerful steering-like inequalities exhibiting better noise tolerance.

As an experimentally relevant case, let us choose $m=5$, in which case the number of communicated bits is $c=\log_2(m-1)=2$. In that case, the LHS(c=2) bound in formula~(\ref{Lc}) becomes $L_2=0.9804$. Due to our result, a Werner state with visibility larger than $L_2$ along with well-chosen measurements violates this two-bit bound $L_2$. In light of recent experimental progress demonstrating EPR steering~\cite{expEPR1,expEPR2,expEPR3}, we believe this bound should be overcome in state-of-the-art photonic experiments.

\section*{Discussion}
In this paper, we extended the notion of Bell inequalities with auxiliary communication to the EPR steering scenario. To do so, we introduced a general framework based on an efficient SDP method. With this tool, we characterized the set of bipartite states which admits a local hidden state model augmented with 1 bit of classical communication (the so-called LHS(1) model) from untrusted Alice to trusted Bob. This $LHS(1)$ set of states was proven to be strictly larger than the set of states admitting an LHS model (for projective measurements). Moreover, this $LHS(1)$ set turns out to have portions outside the $LHV$ set. On the other hand, we conducted an extensive numerical search which indicates that there exist local two-qubit quantum states, which nevertheless cannot be described by an LHS(1) model (assuming projective measurements). We also showed that an infinite amount of classical communication is required from Alice to trusted Bob to simulate the EPR-steering statistics arising from any bipartite pure entangled state.

There is a number of open questions which deserves further investigations.
\begin{itemize}
  \item We found a gap for the visibility $v$ in case of the Werner states between the best LHS(1) model (defining a lower bound) and violation of a steering-like inequality with one bit of communication (defining an upper bound). Would it be possible to close this gap either by improving the lower bound or by improving the upper bound value?
  \item Based on extensive numerical search we conjectured that the $LHV$ set has portions outside the $LHS(1)$ set of states. Is there a formal proof of this conjecture?
  \item We quantified quantum steering with the amount of classical communication between the two parties. What happens if we consider other resources such as certain no-signalling resources?
  \item Another question concerns one-way steerability of quantum states~\cite{oneway}. As an extension of one-way steerable states, we ask whether there exists a bipartite quantum state, such that Alice can steer Bob's state, however, it is impossible for Bob to steer Alice's state even allowing 1 bit of classical communication between them.
  \item It would be also interesting to see how our results relate to LHV models allowing classical communication. We know that 2 bits of communication suffice to simulate projective measurements on any two-qubit entangled state \cite{tonerbacon}. However, in the EPR steering scenario due to our results any finite number of bits is not enough. Does the same result hold true if we add some noise to the singlet state?
  \item In case of $c=1$ we have shown that the nested relation $LHS(c-1)\subsetneq LHS(c)$ holds true. It would be interesting to see if this strict hierarchical relation generalizes to any finite number of $c$ bits.
  \item Finally, it is also interesting to consider the extension of the steering task with communication to the multipartite realm (see, e.g., Refs.~\cite{multist1,multist2,Sainz2015}).
\end{itemize}

\section*{Methods}

\subsection*{Semidefinite program to compute critical weights}\label{appendixA}

Here we provide an SDP program to compute an upper bound on $v_{crit}$ in the formula~(\ref{onepara}).
Assuming the form of the state~(\ref{onepara}), the assemblage $\sigma_{a|x}$ defined by Eq.~(\ref{assem}) in function of parameter $v$ is given by
\begin{equation}
\sigma_{a|x}(v) = v F_{a|x} + (1-v) G_{a|x},
\end{equation}
where
\begin{align}
F_{a|x}&= \Tr_A(|\psi\rangle\langle\psi|M_{a|x}\otimes\one)\nonumber\\
G_{a|x}& = \Tr_A(\rho_{noise}M_{a|x}\otimes\one)
\end{align}
are some fixed matrices. With these expressions in hand, we get the following SDP optimization problem:
\begin{equation} \label{maximsdp}
\begin{aligned}
\text{maximize}\quad & v \\
\text{subject to}\quad & F_{a|x}+(1-v)G_{a|x}=\sum_{\lambda}\sum_{c=1}^d D_{\lambda}(a,c|x)\tilde\sigma_{\lambda,c} &\forall a,x \\
\quad &\Tr \tilde\sigma_{\lambda,c} = \Tr \tilde\sigma_{\lambda,c'} &\forall \lambda,c\neq c'\\
\quad &\sum_{\lambda}\Tr\tilde\sigma_{\lambda,c}=1 &\forall c\\
\quad &\tilde\sigma_{\lambda,c}\ge 0 &\forall \lambda,c
\end{aligned}
\end{equation}

\subsection*{LHS(1) model for a Werner state}\label{appendixB}

Here we present a simulation protocol which gives an LHS model augmented with 1 bit of classical communication for the 2-qubit Werner state up to the visibility $1/\sqrt 2$. We proceed in two steps. Our first protocol will work for visibility up to $2/3$, whereas the second one, building on the first protocol, works up to the higher visibility of $1/\sqrt 2$.


Our first one bit protocol is as follows. Alice and Bob share two independently and uniformly distributed random variables $\vec\lambda_1$ and $\vec\lambda_2$ over the unit sphere. The protocol proceeds as follows:
\begin{enumerate}
\label{protocol1}
  \item Alice receives input vector $\vec a$.
  \item Alice outputs $\alpha = +1$ if $|\vec a\cdot\vec\lambda_0|>|\vec a\cdot\vec\lambda_1|$, otherwise outputs $\alpha=-1$.
  \item Alice sends a bit $c$ to Bob which labels $\vec\lambda_c$, $c=1,2$ for which $\alpha=+1$.
  \item Upon receiving this information, Bob outputs the state $\sigma_{\vec\lambda_c}=\frac{\one+\vec\lambda_c\cdot\vec\sigma}{2}$.
\end{enumerate}

The goal of this protocol is to reproduce the assemblage~(\ref{assem}) originating from a two-qubit Werner state~(\ref{werner}). The assemblage of a Werner state is given by
\begin{equation}
\sigma_{\alpha|\vec a}=\Tr_A(\rho_W(v)M_{\alpha|\vec a}\otimes\one)=\frac{\one+\alpha v\vec a\cdot\vec\sigma}{4},
\end{equation}
where $M_{\alpha|\vec a}=(\one+\alpha\vec a\cdot\sigma)/2 $ are rank-1 projectors, where $\alpha=\pm 1$. Because of redundancy, it is enough to reproduce the following object
\begin{equation}
\label{sigma_werner}
\sigma_{\vec a} = \sigma_{+1|\vec a}-\sigma_{-1|\vec a}=\frac{1}{2}v\vec a\cdot\vec\sigma
\end{equation}
in case of a Werner state with visibility $v$.

On the other hand, the object $\sigma_{\vec a}$ coming from the simulation protocol can be expressed as
\begin{equation}\label{sigma_sim}
\sigma_{\vec a}= \frac{1}{4\pi^2}\int{d\vec\lambda_1d\vec\lambda_2}\times
\begin{cases}
    \textrm{sgn}(\vec a\cdot\vec\lambda_1)\vec\sigma_{\vec\lambda_1}, \text{if}\,|\vec a\cdot\vec\lambda_1|>|\vec a\cdot\vec\lambda_2|\nonumber\\
    \textrm{sgn}(\vec a\cdot\vec\lambda_2)\vec\sigma_{\vec\lambda_2}, \text{otherwise}.
\end{cases}
\end{equation}
Using symmetries, we can further write
\begin{equation}\label{sigma_sim2}
\sigma_{\vec a}= \frac{1}{2}\left(\frac{1}{4\pi^2}\int{d\vec\lambda_1d\vec\lambda_2}\max\{|\vec a\cdot\vec\lambda_1|,|\vec a\cdot\vec\lambda_2|\}\right)\vec a\cdot\vec\sigma.
\end{equation}
Comparing this formula with (\ref{sigma_werner}), the critical visibility $v$ is given by the closed form expression

\begin{equation}\label{sigma_sim3}
v= \frac{1}{4\pi^2}\int{d\vec\lambda_1d\vec\lambda_2}\max\{|\lambda_{1z}|,|\lambda_{2z}|\}=\frac{2}{3}.
\end{equation}

\noindent where we used the fact that because of spherical symmetry we can take $\vec a$ pointing to the north pole (i.e. to positive z-axis), hence $\vec a\cdot\vec\lambda_c=\lambda_{cz}$ for $c=1,2$ and we also used the fact that

\begin{equation}\label{sigma_sim4}
\int{du_1 du_2}\max\{\left(|u_1|,|u_2|\right)\}=\frac{2}{3}.
\end{equation}

\noindent for uniformly distributed $u_1$, $u_2$ in the interval $\left[0,1\right]$.

We now improve the above one bit protocol up to visibility $v=1/\sqrt 2$. To this end, we use the same protocol as before, but this time $\vec\lambda_1$ and $\vec\lambda_2$ are correlated variables. We choose them as $\vec\lambda_1=U\vec e_z$, $\vec\lambda_2=U\vec e_x$, such that the $2\times2$ matrix $U$ is distributed according to the Haar measure on $SU(2)$. In that case, the protocol gives $\sigma_{\vec a}=\sigma_{+1|\vec a}-\sigma_{-1|\vec a}$ with
\begin{equation}\label{sigma_sim5}
\sigma_{\vec a}= \frac{1}{2}\left(\frac{1}{4\pi^2}\int{\nu(U)}\max\{|\vec a\cdot\vec\lambda_1|,|\vec a\cdot\vec\lambda_2|\}\right)\vec a\cdot\vec\sigma,
\end{equation}
\noindent where $\nu(U)$ defines the Haar measure on $SU(2)$ and $\vec\lambda_1=U\vec e_z$, $\vec\lambda_2=U\vec e_x$. Let us set $\vec a \equiv \vec e_z$ by rotating the coordinate system appropriately and denote $u = U\vec e_z$. With these substitutions, we obtain the formula for the critical visibility

\begin{equation}\label{sigma_sim6}
v= \frac{1}{4\pi}\int{d\vec u}\max{\left(|u_z|,|u_x|\right)}=\frac{1}{\sqrt 2},
\end{equation}
where integration was performed over the unit sphere.

\subsection*{The $LHS(1)$ set is strictly larger than the $LHS$ set}\label{AppendixC}
Here we prove the title. For the two-qubit Werner states the $LHS$ set of states is bounded by $v=1/2$~\cite{steering1,steering2,steeringNP}. Hence any LHS(1) model giving a threshold value higher than $v=1/2$ does the job. Hence, the LHS(1) model with threshold $v=1/\sqrt 2$ presented in Methods section previously provides us with the desired proof. We give here a LHS(1) model with a smaller threshold $v=0.5899$. Though, this value is worse than our previous threshold $v=1/\sqrt 2$, the present proof is completely different and maybe of independent interest. In fact, the proof below for an LHS(1) model is a special instance of the algorithmic procedure to construct LHS models appeared in Refs.~\cite{alg1,alg2}.

Let us pick the icosahedron, a platonic solid which has 12 vertices and 20 faces. Using the SDP defined in Methods~A, we compute $v_{crit}=0.7423$ for the measurements pointing toward the 12 vertices of the icosahedron. Note that the icosahedron has a reflection symmetry through the center, and it is enough to take only 6 of its vertices:
\begin{align}
\label{vecu}
\vec u_1 & = (0,1,\varphi)\nonumber\\
\vec u_2 & = (0,1,-\varphi)\nonumber\\
\vec u_3 & = (1,\varphi,0)\nonumber\\
\vec u_4 & = (1,-\varphi,0)\nonumber\\
\vec u_5 & = (\varphi,0,1)\nonumber\\
\vec u_6 & = (-\varphi,0,1)
\end{align}
where $\varphi$ is the golden ratio $\varphi=(1+\sqrt 5)/2$. Following Refs.~\cite{Bowles2015,Sainz2015,alg1,alg2}, any vector $\vec u$ which is within the (largest) inscribed sphere of this icosahedron, can be expressed as the convex combination of the 12 vertices (the ones in (\ref{vecu}) and its inverted versions).  The computation takes roughly 1 min on a normal desktop PC. If we normalize the vertices~(\ref{vecu}) such that all of them have unit length from the origin, the radius of the inscribed sphere is $r=\sqrt{(5 + 2\sqrt 5)/15}\sim0.794654$. Hence, the Werner state with visibility $v_{crit}=0.742344$ has a LHS(1) model for any set of noisy observables of Alice $A = \mu \, A(\vec u)$ for $\mu\le r=0.794654$. As a side remark, we note that the above value of $v_{crit}=0.742344$ can be obtained by using the steering-like inequality $S_m = \sum_{x=1}^m E_{x,x}/m$ presented in the Results section. Indeed, by setting Bob's Bloch vectors in (\ref{naivesteering}) according to (\ref{vecu}) will recover this value up to numerical precision. An optimal LHS(1) strategy is as follows: $r(x)=(1,0,0,0,0,1)$ and $A_x=[1,-1,-1,1,1,1]$. With this strategy, we have to maximize $S_6 = (1/6)(\langle -\vec u_2\cdot\vec\sigma -\vec u_3\cdot\vec\sigma+\vec u_4\cdot\vec\sigma+\vec u_5\cdot\vec\sigma\rangle_{\rho_0}+\langle \vec u_1\cdot\vec\sigma + u_6\cdot\vec\sigma\rangle_{\rho_1})$ over $\rho_0,\rho_1$. The maximum is given by $S_6 = (1/6)(|-\vec u_2-\vec u_3 + \vec u_4 + \vec u_5| + |\vec u_1 + \vec u_6|)=\sqrt{\frac{5}{18}+\frac{11}{18\sqrt 5}}\simeq 0.742344$.

We now use the identity $\Tr_A\left(\mu A\otimes\one\rho_{W}(v)\right) = \Tr_A\left(A\otimes\one\rho_{W}(\mu v)\right)$,
which in words tells us that the statistics of noisy observables $\mu A$ on the Werner state $\rho_{W}(v)$ perfectly match the statistics of noiseless observables $A$ on the Werner state having visibility $\mu v$. We thus have a Werner state with visibility $v\le r v_{crit}\simeq 0.589907$, which gives us a LHS(1) model for $v\le0.5899$ as announced.

\subsection*{No QS($\infty$) model for Werner states for $v>1/2$}\label{AppendixE}
Here we provide a sketch of the proof for the title. We first give an inequality which proves the (known) result that the Werner state is entangled above $v=1/3$. The same inequality will be used to prove that the Werner state does not admit a QS($\infty$) model for $v>1/2$.
The inequality is as follows.
\begin{equation}
\label{ineqQS}
QS_m = \frac{1}{m}\sum_{x,y=1}^m\delta_{x,y}E_{x,y}\le L_c
\end{equation}
which is similar to inequality~(\ref{naivesteering}). Here we assume that both Alice's and Bob's measurements are continuously and evenly distributed on the Bloch sphere, and our task is to compute $L_c$ in case of $c=0$ and $c=\infty$ bits of communication from Alice to Bob.

Let us start with $c=0$. Then we use the definition $E_{x,y}=\sum_{\lambda} P(\lambda)\Tr(\sigma_{\lambda}^A A_x)\Tr(\sigma_{\lambda}^B B_y)$ in the QS model, where $A_x=\vec u_x\cdot\vec\sigma$ and $B_y=\vec u_y\cdot\vec\sigma$. Exploiting spherical symmetry and the convex property of the definition $E_{x,y}$, we can take $\sigma_{\lambda}^A=\sigma_{\lambda}^B=\ketbra{0}{0}$ without loss of generality and the maximization provides
\begin{equation}
\label{czero}
L_0 = \frac{1}{4\pi^2}\int{d\vec u|u_z|^2}=\frac{1}{3},
\end{equation}
where $\vec u$ is distributed uniformly on the unit sphere and $u_z$ denotes $\vec e_z\cdot\vec u$. Note that the maximum of the right-hand-side of inequality~(\ref{ineqQS}) is 1, attainable with a maximally entangled two-qubit state (i.e. Werner state with $v=1$). Then we obtain the result that Werner states with $v>1/3$ violate the quantum separability inequality~(\ref{ineqQS}), hence they are entangled in this range.

Next we deal with $c=\infty$. Since the amount of communication is unbounded, Alice is able to communicate her measurement settings $x$ to Bob, which permits Bob to adjust his hidden state $\sigma_{\lambda}^B$ according to $x$. This in turn implies the maximum
\begin{equation}
\label{cinfty}
L_{\infty} = \max{\int{d\vec u\left|\Tr{(\ketbra{0}{0}\cdot\ketbra{\vec u}{\vec u})}\right|}} = \frac{1}{4\pi^2}\int{d\vec u|u_z|}=\frac{1}{2}.
\end{equation}
Then we obtain the announced result that Werner states with $v>1/2$ violate the quantum separability inequality~(\ref{ineqQS}), therefore there is no QS($\infty$) model for the parameter range $v>1/2$.

\subsection*{Equivalence of states concerning the LHS(c) model}\label{AppendixF}
An LHS(c) model for the 2-qubit Werner state gives rise to the same LHS(c) model for more general quantum states. To this end, we note the recent result on the equivalence of states using local filtering (or more generally of any trace non-increasing CP maps) on Bob's side \cite{marco}. Following the same steps as in the proof of Lemma~2 in Ref.~\cite{marco}, it can be shown that if Bob performs filtering operation on any state which has a LHS(c) model, the resulting state also admits a LHS(c) model. Now let Bob apply a local filter $F_B(\theta)=\cos\theta|0\rangle\langle 0|+\sin\theta|1\rangle\langle 1|$ on the Werner state~(\ref{werner}). The state after this operation becomes $\rho_{\theta}(v) = v|\psi(\theta)\rangle\langle\psi(\theta)|+(1-v)\one/2\otimes\sigma$, and $\sigma=\Tr_A|\psi(\theta)\rangle\langle\psi(\theta)|$, where $|\psi(\theta)\rangle=\sin\theta|01\rangle + \cos\theta|10\rangle$. This result implies that $\rho_{\theta}(v)$ has a LHS(1) model for any $\theta>0$ below $v = 1/\sqrt 2$. However, this threshold may not be tight, that is, it does not rule out the possibility of a higher $v_{crit}(\theta)$ for $\theta\le\pi/4$.

\section*{Acknowledgements}
S.N. and T.V. acknowledge financial support from the Hungarian
National Research Fund OTKA K112233 and K111734, respectively. S.N. was supported by a J\'anos Bolyai Grant of
the Hungarian Academy of Sciences.

\section*{Author contributions statement}

S.N. and T.V. designed and performed the research as well as wrote the paper.

\section*{Competing financial interests}

The authors declare no competing financial interests.

\end{document}